\documentclass[conference]{IEEEtran}
\IEEEoverridecommandlockouts
\usepackage{cite}
\usepackage{amsmath,amssymb,amsfonts}
\usepackage{graphicx}
\usepackage{subcaption}
\usepackage{textcomp}
\usepackage{xcolor}
\usepackage{url}
\usepackage{makecell}
\usepackage{mathtools}
\usepackage{float}
\usepackage{gensymb}
\usepackage{booktabs}
\usepackage{bm}
\usepackage{algorithm}
\usepackage{algpseudocode}
\usepackage[export]{adjustbox}
\def\BibTeX{{\rm B\kern-.05em{\sc i\kern-.025em b}\kern-.08em
    T\kern-.1667em\lower.7ex\hbox{E}\kern-.125emX}}

\usepackage[bottom=0.87in,top=0.7in,left=0.57in,right=0.57in]{geometry}
\usepackage{hyperref}
\hypersetup{
    colorlinks=true,
    linkcolor=blue,
}

\newcommand{\Expect}[1]{\mathbb{E}\{#1\}}
\DeclareMathOperator{\Tr}{Tr}

\begin{document}

% Use et. al. for references with more than three authors:
\bstctlcite{IEEEexample:BSTcontrol}

% \addtolength{\topmargin}{0.36in}

\title{On the Spectral Efficiency of D-MIMO Networks under Rician Fading}

\author{Eduardo Noboro Tominaga\IEEEauthorrefmark{1}, Onel Luis Alcaraz López\IEEEauthorrefmark{1}, Tommy Svensson\IEEEauthorrefmark{2}, Richard Demo Souza\IEEEauthorrefmark{3}, Hirley Alves\IEEEauthorrefmark{1}\\

	\IEEEauthorblockA{
		\IEEEauthorrefmark{1}6G Flagship, Centre for Wireless Communications (CWC), University of Oulu, Finland.\\        
		E-mail: \{eduardo.noborotominaga, onel.alcarazlopez, hirley.alves\}@oulu.fi\\
        \IEEEauthorrefmark{2}Department of Electrical Engineering, Chalmers University of Technology, 412 96 Gothenburg, Sweden.\\ E-mail: tommy.svensson@chalmers.se\\
		\IEEEauthorrefmark{3}Department of Electrical and Electronics Engineering, Federal University of Santa Catarina (UFSC), Florian\'{o}polis,\\88040-370, Brazil. E-mail: richard.demo@ufsc.br\\
	}
}

\maketitle

\begin{abstract}
Contemporary wireless communications systems adopt the Multi-User Multiple-Input Multiple-Output (MU-MIMO) technique: a single base station or Access Point (AP) equipped with multiple antenna elements serves multiple active users simultaneously. Aiming at providing a more uniform wireless coverage, industry and academia have been working towards the evolution from centralized MIMO to Distributed-MIMO. That is, instead of having all the antenna elements co-located at a single AP, multiple APs, each equipped with a few or a single antenna element, jointly cooperate to serve the active users in the coverage area. In this work, we evaluate the performance of different D-MIMO setups under Rician fading, and considering different receive combining schemes. Note that the Rician fading model is convenient for MU-MIMO performance assessment, as it encompasses a wide variety of scenarios. Our numerical results show that the correlation among the channel vectors of different users increases with the Rician factor, which leads to a reduction on the achievable Spectral Efficiency (SE). Moreover, given a total number of antenna elements, there is an optimal number of APs and antenna elements per AP that provides the best performance. This ``sweet spot'' depends on the Rician factor and on the adopted receive combining scheme.
\end{abstract}


\section{Introduction}

\par Multi-User Multiple-Input Multiple-Output (MU-MIMO) technologies are instrumental in enhancing the performance of wireless communication networks. In MU-MIMO networks, a base station or Access Point (AP) equipped with multiple antenna elements serves multiple devices simultaneously. Relying on beamforming techniques, MIMO offers several benefits such as diversity and array gains, spatial multiplexing and interference suppression capabilities that collectively enhance the capacity, reliability, and coverage of wireless systems \cite{heath2018}.

\par Current MU-MIMO systems adopt a centralized paradigm: a coverage area is served by a single base station or Access Point (AP) equipped with multiple antenna elements. The main drawback of the centralized MIMO approach is the significant performance gap between the users located close to the AP and the users located far away from the AP. Aiming at solving this issue, industry and academia have been working towards the evolution from centralized MU-MIMO to Distributed MIMO (D-MIMO) networks. That is, instead of deploying a single AP equipped with many antenna elements, a coverage area is served by multiple APs, each equipped with a few or a single antenna element. All APs are connected to a common Central Processing Unit (CPU), which is responsible for performing the signal processing and coordination tasks \cite{interdonato2019}.

\par Understanding how a MU-MIMO system performs under different conditions helps optimize the system for higher data rates, higher reliability and lower latency. One of the optimization problems associated with D-MIMO systems involves resource allocation, that is, determining the number of APs and the number of antenna elements per AP. This allocation directly impacts the deployment and maintenance costs, the energy efficiency and the scalability of the system.

% Related works:

\par \textbf{Related works:} The Rician fading channel model has been adopted in several works to numerically evaluate the performance of communication systems. This channel model encompasses a broad range of scenarios, since it takes into account a deterministic Line-of-Sight (LoS) component and a random Non-LoS (NLoS) component. The ratio of the signal power in the LoS component to the NLoS component is referred to as the Rician factor. However, when evaluating the impact of the Rician factor on the achievable Spectral Efficiency (SE) of MU-MIMO systems, different works present contradictory results. For instance, the works \cite{boukhedimi2019,kassaw2018,ghacham2017} evaluated the uplink performance of a single-cell MU-MIMO system, and showed that the SE increases with the Rician factor. In contrast, other works showed that the Rician factor can be detrimental to the SE. In the case of a single user MIMO systems, the authors in \cite{zhang2016} showed that the rank of the channel matrix decreases with the Rician factor, and as a consequence the achievable SE also decreases. In the case of MU-MIMO systems where the users are equipped with multiple antennas, the achievable SE also decreases with the Rician factor \cite{berhane2017,he2021}.  In \cite{ghacham2019}, the authors considered a single cell MU-MIMO system with imperfect hardware, and showed that the SE decreases with the Rician factor. Interestingly, some works showed than the Rician factor can be either beneficial or detrimental for the achievable SE. Zhang et. al. \cite{zhang2014} evaluated the uplink performance of a single-cell MU-MIMO system considering both Maximum Ratio Combining (MRC) and Zero Forcing (ZF). Their results show that the uplink sum rate increases with the Rician factor in the case of MRC and decreases in the case of ZF. Kammoun et. al. \cite{kammoun2019} evaluated the downlink of a single-cell MU-MIMO system, and showed that the per-user average SE can either increase or decrease with the Rician factor, depending on the spatial distribution of the users.

% \par Considering D-MIMO systems with a given total number of antenna elements, some works studied the trade-off between the number of APs and antenna elements per AP, for instance, \cite{tominaga2024}. Nevertheless, they did not evaluate the ``sweet-spot'' as a function of the Rician factor.

% Contributions:

\par \textbf{Contributions:} In this work, we evaluate the uplink performance of different D-MIMO setups in terms of the mean per-user achievable Spectral Efficiency (SE). Given a total number of antenna elements, we evaluate the trade-off between the number of APs and the number of antenna elements per AP. We adopt the Rician channel model, and evaluate the performance of different setups as a function of the Rician factor. Our numerical results show that, when the Rician factor increases, the correlation among the channel vectors also increases, which leads to a reduction of the achievable SE. % This reduction of SE with the Rician factor is observed when ZF is used and also in the case of D-MIMO with MRC.
Moreover, there is an optimal number of APs (and antenna elements per AP) that achieves the best performance, which evinces the existence of a trade-off between beamforming and macro-diversity gains. This ``sweet spot'' also depends on the adopted receive combining scheme.

% Organization of the paper:

\par This paper is organized as follows. The considered system model is introduced in Section \ref{systemModel}. We define metrics for the correlation among the channel vectors in Section \ref{linearCorrelation}. Numerical results and discussions are presented in Section \ref{numericalResults}. Finally, we draw the conclusions of this work in Section \ref{conclusions}.

\par \textbf{Notation:} lowercase bold face letters denote column vectors, while boldface upper case letters denote matrices. $a_i$ is the $i$-th element of the column vector $\textbf{a}$, while $\textbf{a}_i$ is the $i$-th column of the matrix $\textbf{A}$.  $[A]_{i,j}$ is the $i$-th row, $j$-th column element of the matrix $\textbf{A}$.
$\textbf{I}_M$ is the identity matrix with size $M\times M$. The superscripts $(\cdot)^T$ and $(\cdot)^H$ denote the transpose and the conjugate transpose operators, respectively. The magnitude of a scalar quantity or the cardinality of a set are denoted by $|\cdot|$. The Euclidian norm of a vector (2-norm) is denoted by $\Vert\cdot\rVert$. We denote the one dimensional uniform distribution with bounds $a$ and $b$ by $\mathcal{U}(a,b)$. We denote the multivariate Gaussian distribution with mean $\mathbf{a}$ and covariance $\mathbf{B}$ by $\mathcal{N}(\mathbf{a},\mathbf{B})$.

\section{System Model}
\label{systemModel}

\par
\begin{figure}
    \centering
    \includegraphics[scale=0.5]{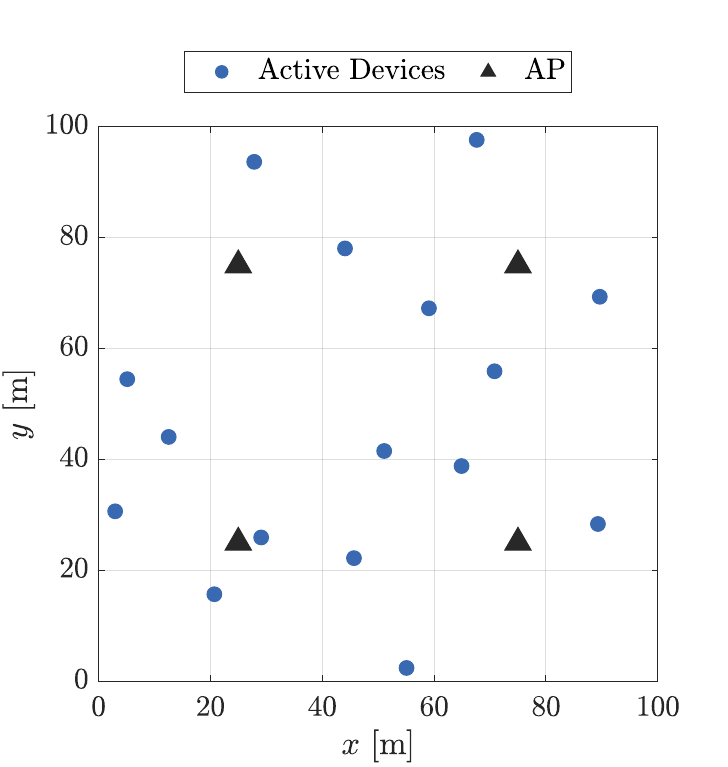}
    \caption{Illustration of the considered system model for $Q=4$, $l=100$ m and $K=16$.}
    \label{illustrationSystemModel}
\end{figure}

 We consider a square area with dimensions $l\times l\;\text{m}^2$, wherein $K$ single-antenna devices are served jointly by $Q$ APs. Each AP is placed at height $h_{\text{AP}}$ and is equipped with a Uniform Linear Array (ULA) of $S$ half-wavelength spaced antenna elements. 
Let $\textbf{p}_k=(x_k,y_k)^T$ denote the coordinates of the $k$-th device, assuming for simplicity that all devices are positioned at the same height $h_{\text{UE}}$. The system model is illustrated in Fig. \ref{illustrationSystemModel} and we focus on the uplink.

\subsection{Channel Model}

\par We adopt a spatially correlated Rician fading channel model \cite{ozdogan2019}. Let $\textbf{h}_{kq}\in\mathbb{C}^{S\times1}$ denote the channel vector between the $k$-th device and the $q$-th AP. It can be modeled as \cite{dileep2021}
\begin{equation}
    \label{channelVector}
    \textbf{h}_{kq}=\sqrt{\dfrac{\kappa}{1+\kappa}}\textbf{h}_{kq}^{\text{los}} + \sqrt{\dfrac{1}{1+\kappa}}\textbf{h}_{kq}^{\text{nlos}},
\end{equation}
where $\kappa$ is the Rician factor, $\textbf{h}_{kq}^{\text{los}}\in\mathbb{C}^{S\times1}$ is the deterministic LoS component, and $\textbf{h}_{kq}^{\text{nlos}}\in\mathbb{C}^{S\times1}$ is the random NLoS component.

\par The deterministic LoS component is given by
\begin{equation}
    \label{losComponent}
    \textbf{h}_{kq}^{\text{los}}=\sqrt{\beta_{kq}}
    \begin{bmatrix}
        1\\
        \exp(-j2\pi\Delta\sin(\phi_{kq}))\\
        \exp(-j4\pi\Delta\sin(\phi_{kq}))\\
        \vdots\\
        \exp(-j2\pi(S-1)\Delta\sin(\phi_{kq}))\\
    \end{bmatrix},
\end{equation}
where $\beta_{kq}$ is the power attenuation owing to the distance between the $k$-th device and the $q$-th AP, $\Delta$ is the normalized inter-antenna spacing, and $\phi_{kq}\in[0,2\pi]$ is the azimuth angle relative to the boresight of the ULA of the $q$-th AP. Meanwhile, the random NLoS component is distributed as
\begin{equation}
    \textbf{h}_{kq}^{\text{nlos}}\sim\mathcal{CN}(\textbf{0},\textbf{R}_{kq}).
\end{equation}
Note that
\begin{equation}
    \textbf{h}_{kq}\sim\mathcal{CN}\left(\sqrt{\dfrac{\kappa}{1+\kappa}}\textbf{h}_{kq}^{\text{los}},\dfrac{\textbf{R}_{kq}}{\kappa+1}\right),
\end{equation}
where $\textbf{R}_{kq}\in\mathbb{C}^{S\times S}$ is the positive semi-definite covariance matrix describing the spatial correlation of the NLoS components. Considering uncorrelated fading, the spatial covariance matrix is simply the identity matrix, that is, $\textbf{R}_{kq}=\textbf{I}_S$.

\par The estimated channel vector of the $k$-th device, $\Hat{\textbf{h}}_{kq}\in\mathbb{C}^{S\times1}$, can be modeled as the sum of the true channel vector plus a random error vector as \cite{wang2012,eraslan2013}
\begin{equation}
    \hat{\textbf{h}}_{kq}=\textbf{h}_{kq}+\Tilde{\textbf{h}}_{kq},
\end{equation}
where $\Tilde{\textbf{h}}_{kq}\sim\mathcal{CN}(\textbf{0},\sigma_{\text{csi}}^2\textbf{I})$ is the vector of channel estimation errors. Assuming orthogonal pilot sequences during the uplink data transmission phase and least squares channel estimation, which does not exploit knowledge of the channel statistics, the true channel realizations and the channel estimation errors are not correlated \cite{onel2022}.
The parameter $\sigma_{\text{csi}}^2$ indicates the quality of the channel estimates, and can be modeled as\cite{wang2012,eraslan2013,onel2022}
\begin{equation}
    \sigma_{\text{csi}}^2=\dfrac{1}{K\rho},
\end{equation}
where $\rho$ is the transmit signal-to-noise ratio.

\subsection{Signal Model}

\par The matrix $\textbf{H}\in\mathbb{C}^{M\times K}$ containing the channel vectors between the $K$ devices and the $Q$ APs can be written as
\begin{equation}
    \label{channelMatrix}
    \textbf{H}=[\textbf{h}_1,\textbf{h}_2,\ldots,\textbf{h}_K],
\end{equation}
where $\textbf{h}_k=[\textbf{h}_{k1}^T,\ldots,\textbf{h}_{kQ}^T]^T$, $k\in\{1,\ldots,K\}$. 
Then, the $M\times 1$ received signal vector can be written as
\begin{equation}
    \textbf{y}=\sqrt{p}\textbf{H}\textbf{x}+\textbf{n},
\end{equation}
where $p$ is the fixed uplink transmit power that is the same for all devices, $\textbf{x}\in\mathbb{C}^{K\times 1}$ is the vector of symbols simultaneously transmitted by the $K$ devices, and $\textbf{n}\in\mathbb{C}^{M\times 1}$ is the vector of additive white Gaussian noise samples such that $\textbf{n}\sim\mathcal{CN}(\textbf{0}_{M\times1},\sigma^2_n\textbf{I}_M)$. Note that $\rho=p/\sigma_n^2$.

\par Let $\textbf{V}\in\mathbb{C}^{M\times K}$ be a linear detector matrix used for the joint decoding of the signals transmitted from the $K$ devices. The received signal after the linear detection operation is split in $K$ streams and given by
\begin{equation}
    \textbf{r}=\textbf{V}^H\textbf{y}=\sqrt{p}\textbf{V}^H\textbf{H}\textbf{x}+\textbf{V}^H\textbf{n}.
\end{equation}
Let $r_k$ and $x_k$ denote the $k$-th elements of $\textbf{r}$ and $\textbf{x}$, respectively. Then, the received signal corresponding to the $k$-th device can be written as
\begin{equation}
    \label{r_k}
    r_k=\underbrace{\sqrt{p}\textbf{v}_k^H\textbf{h}_kx_k}_\text{Desired signal} + \underbrace{\sqrt{p}\textbf{v}_k^H\sum_{k'\neq k}^K \textbf{h}_{k'}x_{k'}}_\text{Inter-user interference} + \underbrace{\textbf{v}_k^H\textbf{n}}_\text{Noise},
\end{equation}
where $\textbf{v}_k$ and $\textbf{h}_k$ are the $k$-th columns of the matrices $\textbf{V}$ and $\textbf{H}$, respectively. From (\ref{r_k}), the signal-to-interference-plus-noise ratio of the uplink transmission from the $k$-th device is given by
\begin{equation}
    \label{gamma_k}
    \gamma_k=\dfrac{p|\textbf{v}_k^H\textbf{h}_k|^2}{p\sum_{k'\neq k}^K |\textbf{v}_k^H\textbf{h}_{k'}|^2+\sigma^2_n\lVert\textbf{v}_k^H\rVert^2}.
\end{equation}

\par The receive combining matrix $\textbf{V}$ is computed as a function of the matrix of estimated channel vectors $\hat{\textbf{H}}\in\mathbb{C}^{M\times K}$, $\hat{\textbf{H}}=[\hat{\textbf{h}}_1,\ldots,\hat{\textbf{h}}_K]$. In this work, we compare three different linear receive combining schemes: MRC, ZF and MMSE. For each scheme, the receive combining matrix is computed as \cite{liu2016_2}
\begin{equation}
    \label{precodingMatrices}
    \textbf{V}=
    \begin{cases}
        \hat{\textbf{H}},&\text{ for MRC,}\\
        \hat{\textbf{H}}(\hat{\textbf{H}}^H\hat{\textbf{H}})^{-1},&\text{ for ZF,}\\
        \left(\hat{\textbf{H}}\hat{\textbf{H}}^H+\dfrac{\sigma_n^2}{p}\textbf{I}_M\right)^{-1}\hat{\textbf{H}},&\text{ for MMSE.}
    \end{cases}
\end{equation}

\subsection{Performance Metrics}

\par We adopt as the performance metric the per-user mean achievable uplink SE. The achievable uplink SE of the $k$-th device is \cite{liu2020}
\begin{equation}
    \label{per-user-achievable-SE}
    R_k=\mathbb{E}_{\textbf{H}}\{\log_2(1+\gamma_k)\}.
\end{equation}

\par Then, the per-user mean achievable uplink SE is obtained by averaging over the achievable uplink SE of the $K$ devices,~i.e.,
\begin{equation}
    \label{mean-per-user-achievable-SE}
    \Bar{R}=\dfrac{1}{K}\sum_{k=1}^K R_k.
\end{equation}

\section{Linear Correlation Between Random Vectors}
\label{linearCorrelation}

\par In this section, we define numerical metrics for the correlation among the channel vectors.

\par Let $\textbf{h}_1,\textbf{h}_2\in\mathbb{C}^{M\times1}$ be two random vectors with correlation matrices $\textbf{C}_1,\textbf{C}_2\in\mathbb{C}^{M\times M}$, $\textbf{C}_i=\Expect{\textbf{h}_i\textbf{h}_i^H},\;i\in\{1,2\}$. The correlation coefficient is defined as \cite{puccetti2022}
\begin{equation}
    \label{corrCoeff}
    r(\textbf{h}_1,\textbf{h}_2)=\dfrac{\Tr(\textbf{C}_{12})}{\Tr\left((\textbf{C}_1\textbf{C}_2)^{1/2}\right)},
\end{equation}
where $\textbf{C}_{12}=\Expect{\textbf{h}_1\textbf{h}_2^H}$ is the cross-correlation matrix. Note that (\ref{corrCoeff}) is a generalization of the Pearson coefficient for multivariate random variables, thus $\rho\in[-1,1]$.

\par Finally, we define the average correlation coefficient as the expected absolute value of the correlation coefficient, that is,
\begin{equation}
    \Bar{r}\triangleq\mathbb{E}_{\textbf{h}_1,\textbf{h}_2}\{|r|\},
\end{equation}
which is obtained by averaging over several realizations of the random vectors $\textbf{h}_1$ and $\textbf{h}_2$.

\par An alternative way to illustrate the correlation among channel vectors is by evaluating the mean condition number of the channel matrix. The condition number is a standard measure of how ill-conditioned a matrix is. The condition number $C$ of a matrix $\textbf{H}$ is calculated as
\begin{equation}
    C(\textbf{H})=\dfrac{\sigma_{\text{max}}}{\sigma_{\text{min}}}\geq1,
\end{equation}
where $\sigma_{\text{max}}$ and $\sigma_{\text{min}}$ are the largest and the smallest singular value of the matrix $\textbf{H}$, respectively.

\section{Numerical Results}
\label{numericalResults}

\par In this section, we present our Monte Carlo simulation results. We set the total number of antenna elements $M$, and evaluate the trade-off between the number of APs and antenna elements per AP by varying $Q$ and $S=M/Q$ accordingly. The mean per-user achievable SE is obtained by averaging over a large number of network realizations, i.e., set of positions of the $K$ devices. For each network realization, the results are averaged over several channel realizations, i.e., realizations of the channel matrix $\textbf{H}$.

\subsection{Simulation Parameters}

\par The power attenuation due to the distance (in dB) is modelled using the log-distance path loss model and given by
\begin{equation}
    \beta_{kq}=-L_0-10\eta\log_{10}\left(\dfrac{d_{kq}}{d_0}\right),
\end{equation}
where $d_0$ is the reference distance in meters, $L_0$ is the attenuation owing to the distance at the reference distance (in dB), $\eta$ is the path loss exponent and $d_{kq}$ is the distance between the $k$-th device and the $q$-th AP in meters.

\par The attenuation at the reference distance is calculated using the Friis free-space path loss model and given by
\begin{equation}
    L_0=20\log_{10}\left(\dfrac{4\pi d_0}{\lambda}\right),
\end{equation}
where $\lambda$ is the wavelength in meters, $c$ is the speed of light and $f_c$ is the carrier frequency.

\par Unless stated otherwise, the values of the simulation parameters are listed on Table \ref{tableParameters}. The $K$ active MTDs are uniformly distributed on the square coverage area, that is, $x_k,y_k\sim\mathcal{U}[0,l]$. Considering the selected values of $M$ and $h_{\text{AP}}$, the communication links between the AP and any device experience far-field propagation conditions. The noise power (in Watts) is given by $\sigma^2_n=N_0BN_F$, where $N_0$ is the power spectral density of the thermal noise in W/Hz, $B$ is the signal bandwidth in Hz, and $N_F$ is the noise figure at the receivers.

\begin{table}[t]
    \centering
    \caption{Simulation parameters.}
    \begin{tabular}{l l l}
        \toprule
        \textbf{Parameter} & \textbf{Symbol} & \textbf{Value}\\  
        \midrule
        Total number of antenna elements & $M$ & $16$\\
        Number of APs & $Q$ & $[1,2,4,8,16]$\\
        Number of antenna elements per AP & $S$ & $[16,8,4,2,1]$\\
        Number of active devices & $K$ & $16$\\
        Length of the side of the square area & $l$ & 100 m\\
        Uplink transmission power & $p$ & 100 mW\\
        PSD of the noise & $N_0$ & $4\cdot10^{-21}$ W/Hz\\
        Signal bandwidth & $B$ & 20 MHz\\
        Noise figure & $N_F$ & 9 dB\\
        Height of the APs & $h_{\text{AP}}$ & 12 m\\
        Height of the UEs & $h_{\text{UE}}$ & 1.5 m\\
        Carrier frequency & $f_c$ & 3.5 GHz\\
        Normalized inter-antenna spacing & $\Delta$ & $0.5$\\
        Path loss exponent & $\eta$ & $2$\\
        Reference distance & $d_0$ & $1$ m\\
        \bottomrule
    \end{tabular}    
    \label{tableParameters}
\end{table}

\subsection{Correlation Analysis}

\begin{figure}[t]
    \centering
    \includegraphics[scale=0.55]{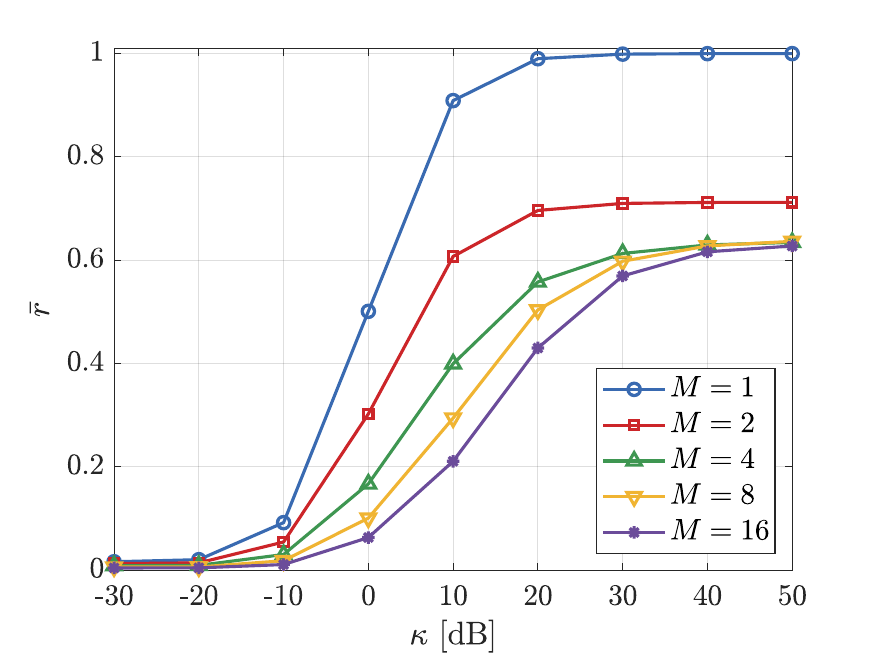}
    \caption{Average correlation coefficient between two channel vectors as a function of the Rician factor, for a single AP with varying number of antennas.}
    \label{plotCorrelationCoefficient}
    ~
    \centering
    \includegraphics[scale=0.55]{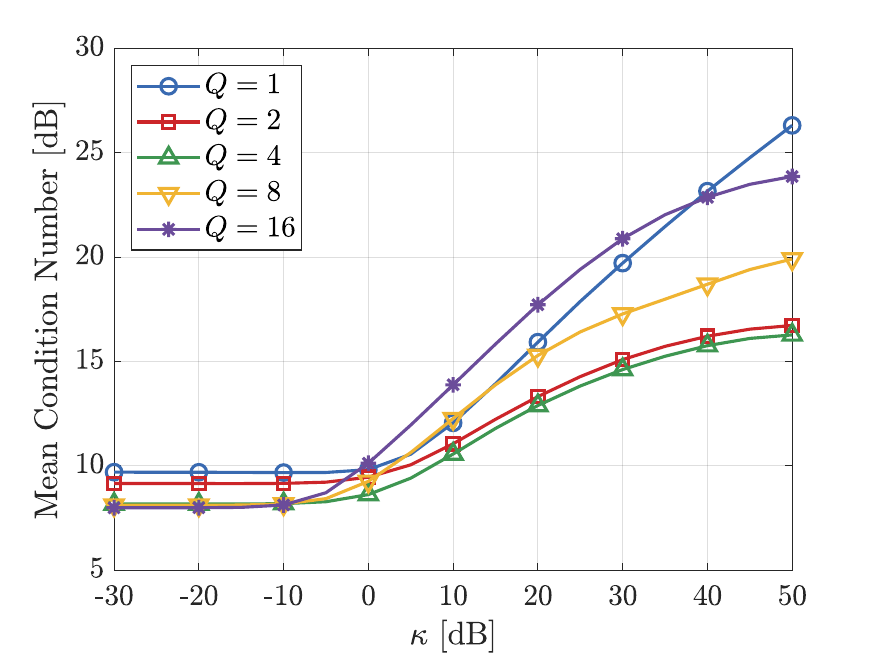}
    \caption{Mean condition number of the channel matrix $\textbf{H}$ versus the Rician factor, for $M=16$, $K=10$ and $l=100$~m.}
    \label{plotCondNumber}
\end{figure}

\par We first show that the correlation among channel vectors increases with the Rician factor $\kappa$, which yields a reduction on the mean per-user achievable SE.

\par Fig. \ref{plotCorrelationCoefficient} shows the average correlation coefficient $\Bar{\rho}$ of the channel vectors of two different devices versus the Rician factor $\kappa$, considering $Q=1$, different numbers of antenna elements at the AP. We observe that the correlation between the channel vectors increases with Rician factor. When $\kappa\rightarrow0$, which corresponds to a purely NLoS channel, the channel vectors are uncorrelated. Then, when $\kappa$ grows large, the correlation significantly increases. Nevertheless, the correlation coefficient grows slower as we increase the number of antenna elements at the AP. Increasing the number of antennas at the AP enhances its spatial multiplexing capabilities, thus the AP becomes better at separating the users in the spatial domain.

\begin{figure*}[!t]
    \centering
    \subfloat[$Q=1$]{\includegraphics[width=0.2\textwidth]{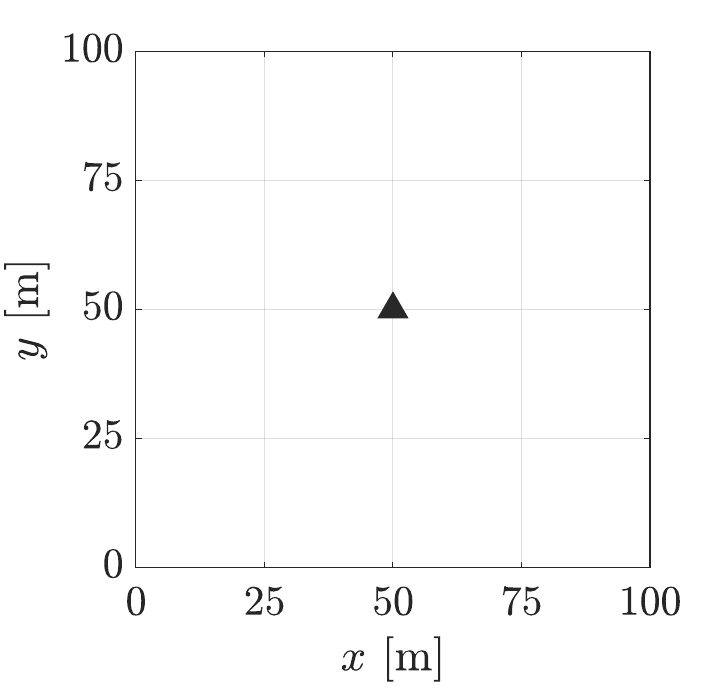}\label{fig1}} \hfill
    \subfloat[$Q=2$]{\includegraphics[width=0.2\textwidth]{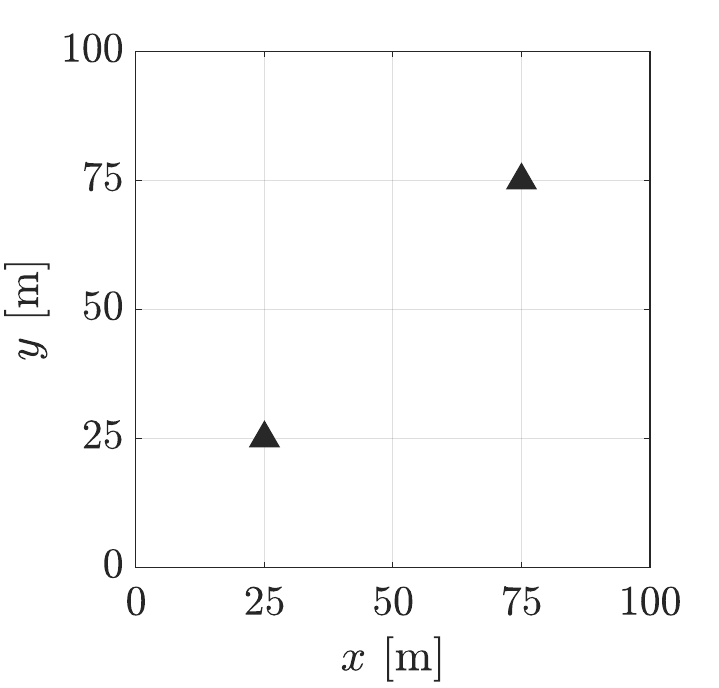}\label{fig2}} \hfill
    \subfloat[$Q=4$]{\includegraphics[width=0.2\textwidth]{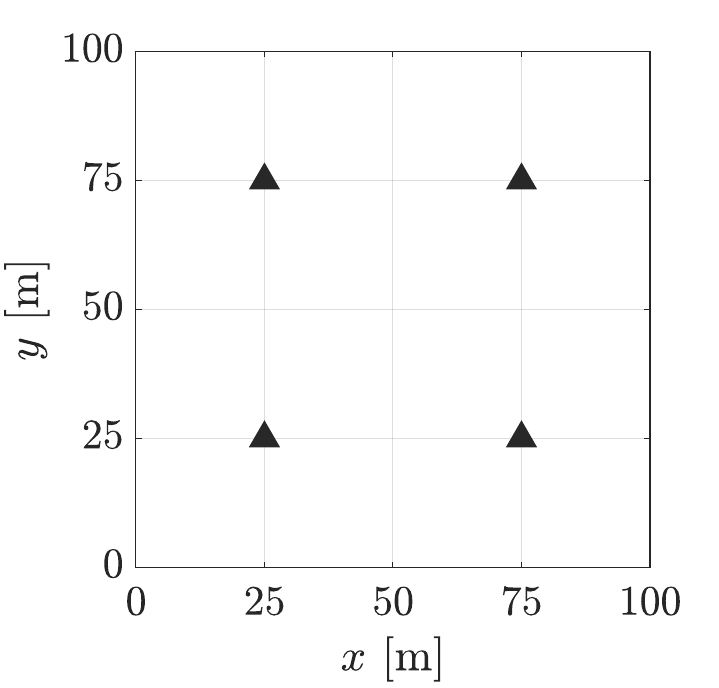}\label{fig3}}
    \subfloat[$Q=8$]{\includegraphics[width=0.2\textwidth]{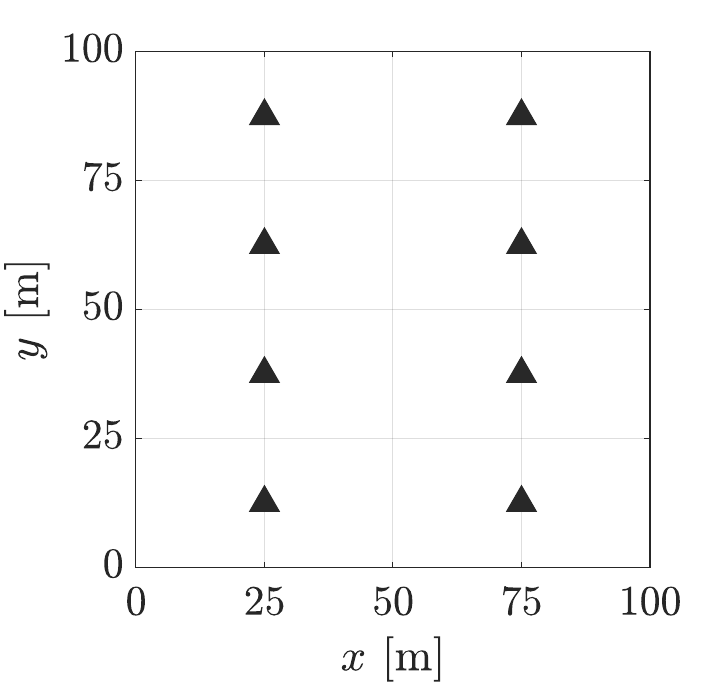}\label{fig4}} 
    \subfloat[$Q=16$]{\includegraphics[width=0.2\textwidth]{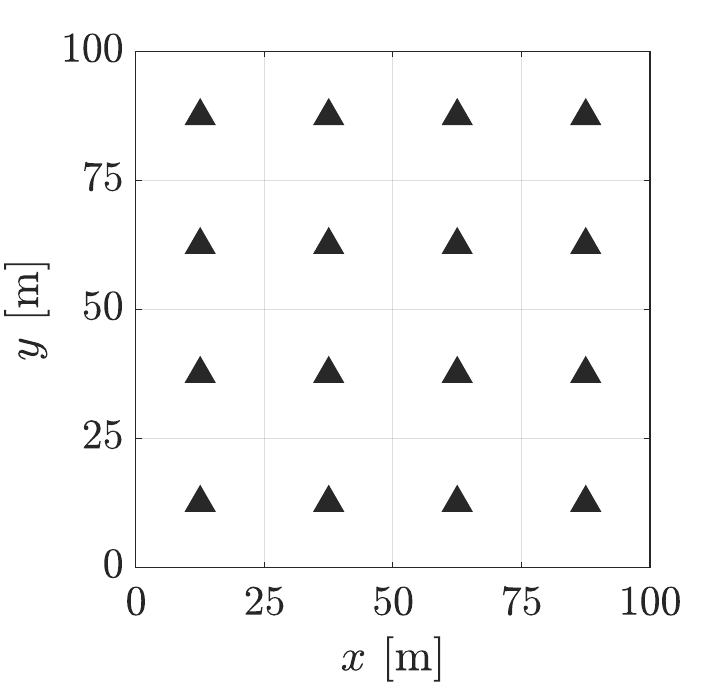}\label{fig5}}
    \caption{Illustrations of the different D-MIMO setups considered in this work.}
    \label{setup_APs}
\end{figure*}

\par We numerically compute\footnote{The condition number of a matrix can be easily computed in Matlab using the function \texttt{cond}. By default, this function computes the 2-norm condition number \cite{mathworks_cond}.} the mean condition number $\Bar{C}$ of the channel matrix $\textbf{H}$, which was defined in (\ref{channelMatrix}). The mean condition number is obtained by averaging over multiple network realizations, that is, multiple set of positions of the $K$ devices. For each network realization, the results are averaged over several channel realizations.

\par Fig. \ref{plotCondNumber} shows the mean condition number\footnote{Note that some RF measurement devices measure the condition number of MIMO channel matrices using the dB scale \cite{rohde2011}.} of the channel matrix $\textbf{H}$ versus the Rician factor. We set the total number of antenna elements $M$, and evaluate the condition number considering different number of antenna elements per AP, $Q$. The curves show that, for any value of $Q$ and for both the considered types of fading, the mean condition number grows rapidly when $\kappa\geq0$~dB, that is, when the power of the LoS component of the channel vector grows.

\subsection{Results and Discussions}

\par
\begin{figure}[!h]
    \centering
    \subfloat[]{\includegraphics[width=0.45\textwidth]{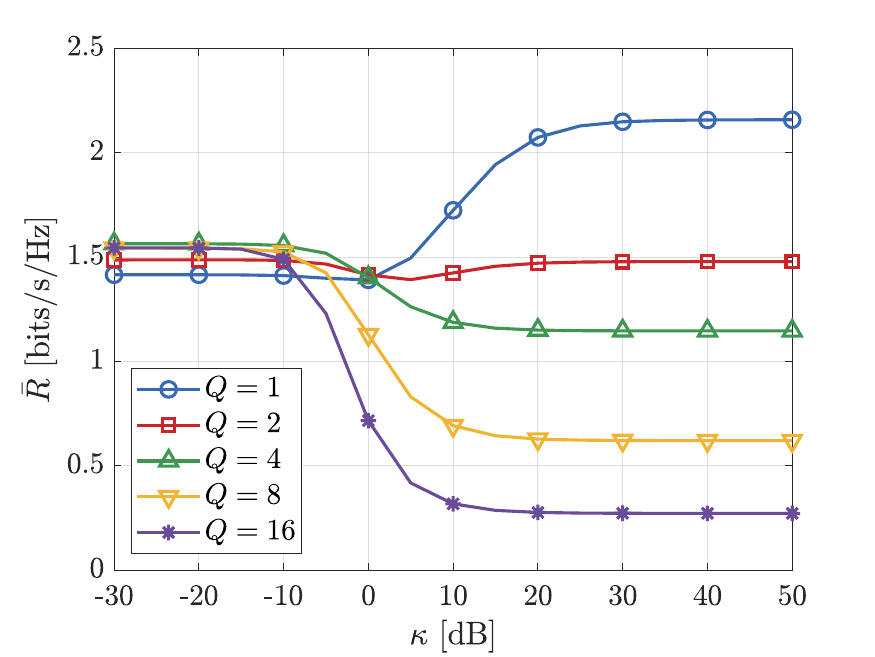}\label{results_SE_MRC}} \\
    \subfloat[]{\includegraphics[width=0.45\textwidth]{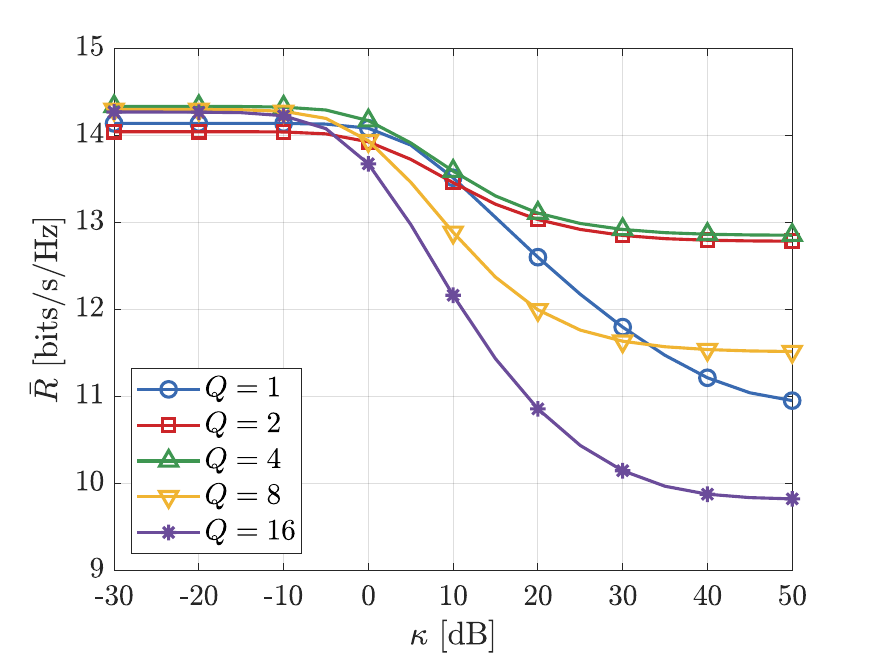}\label{results_SE_ZF}} \\
    \subfloat[]{\includegraphics[width=0.45\textwidth]{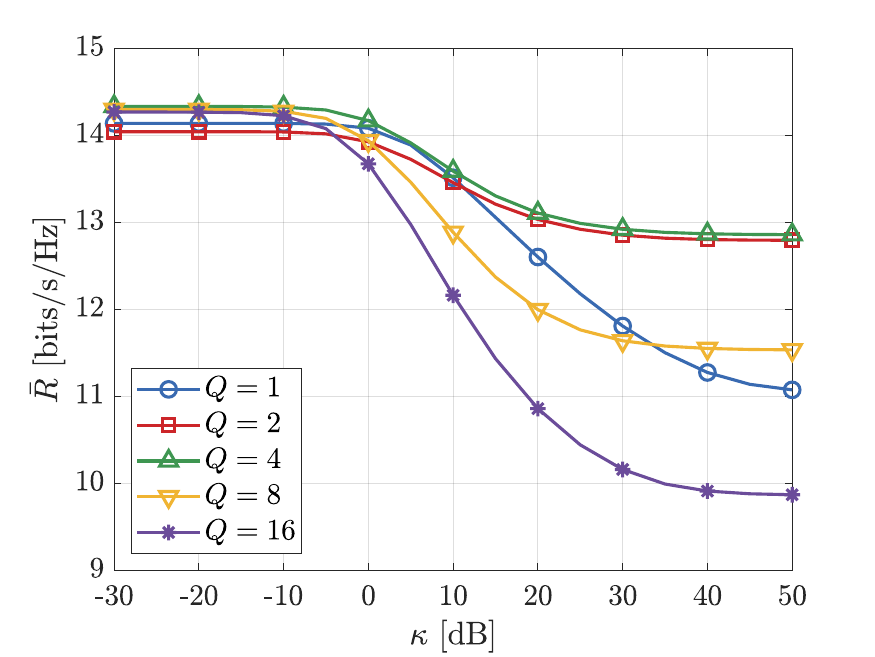}\label{results_SE_MMSE}}
    \caption{Mean per-user achievable SE versus the Rician factor considering (a) MRC (b) ZF and (c) MMSE.}
    \label{results_SE}
\end{figure}

 In Fig. \ref{results_SE}, we plot the mean per-user achievable SE versus the Rician factor considering MRC, ZF, and MMSE. In the case of MRC, the mean per-user achievable SE $\Bar{R}$ increases with the Rician factor $\kappa$ for the case of $Q=1$, and decreases for any other values of $Q$. On the other hand, when ZF or MMSE is adopted, $\Bar{R}$ always decreases with $\kappa$, independently of $Q$. As we increase $\kappa$, the correlation between the channel vectors of the devices increases, which leads to the reduction of the achievable SE. Note that, for the single AP case ($Q=1$), our numerical results are aligned with the results reported in \cite{zhang2014}: the achievable SE increases with $\kappa$ for MRC, while it decreases with $\kappa$ for ZF.

\par When MRC is adopted (Fig. \ref{results_SE_MRC}), we observe a remarkable shift of the ``sweet spot" of $(Q,S)$ as we vary $\kappa$: the best performance is achieved with most distributed setups ($Q=\{4,8,16\}$) when the channel is NLoS dominant ($\kappa\leq -10$ dB) and with the fully centralized setup ($Q=1$) when the channel is LoS dominant ($\kappa\geq10$ dB). Nevertheless, note that the setups with $Q=\{4,8,16\}$ present almost the same performance when $\kappa\leq -10$ dB. In summary, the performance increases with $Q$ in predominantly NLoS channels, whereas it severely decreases with $Q$ in predominantly LoS channels.

\par Conversely, when ZF or MMSE is adopted (Figs. \ref{results_SE_ZF} and \ref{results_SE_MMSE}), we observe again that the most distributed setups  achieve the best performance when the channel is NLoS dominant. The setups with $Q=\{4,8,16\}$ perform very similarly when $\kappa\leq -10$ dB . However, as we increase $\kappa$, the partially distributed setups with $Q=\{2,4\}$ emerge as the best options. When $\kappa$ is very large, we interestingly observe that the most distributed setup with $Q=16$ presents the worst performance, followed by the fully centralized setup with $Q=1$. These results evince the existence of a sweet-spot between the beamforming gains obtained with multi-antenna APs and macro-diversity gains obtained by the spatial distribution of APs, and are aligned with the results from \cite{tominaga2024}.

\par The numerical results shown in Fig. \ref{results_SE} demonstrate that ZF and MMSE combining significantly outperform MRC in terms of mean per-user achievable SE. Nevertheless, they come at the cost of higher computational complexity, since both schemes require matrix inversions for the computation of the receive combining matrix. Besides, the results show that the partially distributed setups with $Q=\{2,4\}$ are the best deployment options. Adopting ZF or MMSE, they strike a balance between the beamforming and macro-diversity gains over the whole range of $\kappa$. Additionally, they reduce the deployment and maintenance cost of the D-MIMO system, since they require the installation of less pieces of equipment and fronthaul connections.

\par
\begin{figure}[!h]
    \centering
    \includegraphics[scale=0.55]{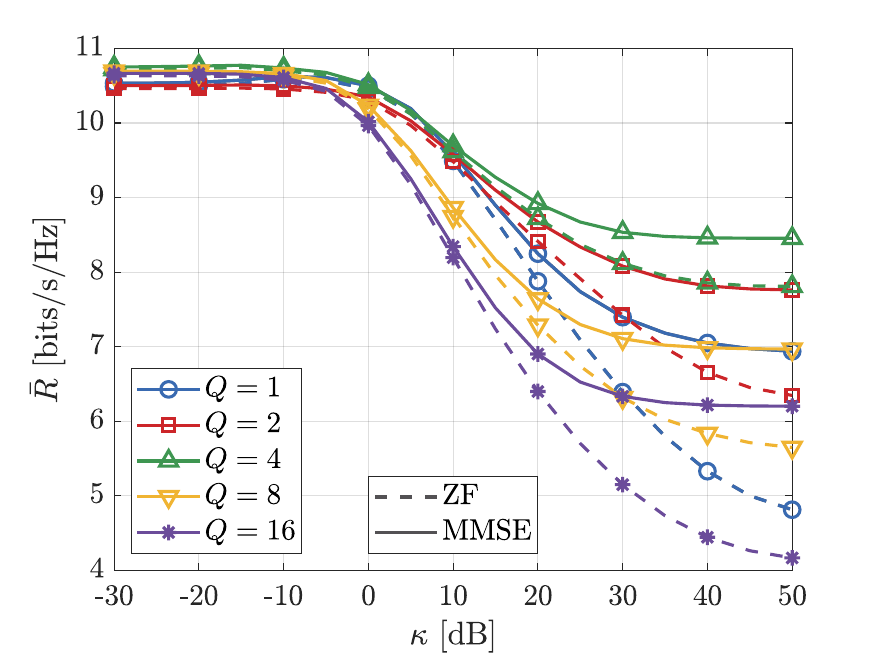}
    \caption{Mean per-user achievable SE versus the Rician factor considering ZF and MMSE, for $M=K=16$.}
    \label{results_ZF_MMSE}
\end{figure}

 The curves in Figs. \ref{results_SE_ZF} and \ref{results_SE_MMSE} show that, in the high SNR regime, the performance of ZF matches that of MMSE. Similarly to those curves, Fig. \ref{results_ZF_MMSE} compares the performance of ZF and MMSE for $K=M=16$, which corresponds to a high inter-user interference scenario. In this case, we can observe clearly that MMSE outperforms ZF when $\kappa$ grows large. Note that our results are consistent with those reported in \cite{liu2016_2}. In their study, the authors analyzed the uplink performance of MU-MIMO systems by considering outage probabilities and compared the performance of MRC, ZF, and MMSE. They observed that ZF and MMSE exhibit comparable performance when $M>K$, whereas MMSE notably surpasses ZF when $M=K$.

\section{Conclusions}
\label{conclusions}

\par In this work, we studied the performance of D-MIMO networks under Rician fading in an indoor scenario. The performance was evaluated in terms of the mean per-user achievable SE in the uplink. We set a total number of antenna elements, and investigated the trade-off between the number of APs and number of antenna elements per AP. We also compared the performance achieved with MRC, ZF and MMSE. Our numerical results show that the correlation between the channel vectors increases with the Rician factor, which yields a reduction on the achievable SE. In addition, the performance of ZF and MMSE significantly exceeds that of MRC. In a nutshell, when the channels are dominantly NLoS, the best performance is achieved by the most distributed D-MIMO setups, i.e., with many APs and few antenna elements per AP. On the other hand, when the channels are dominantly LoS, it is more advantageous to have fewer APs equipped with more antenna elements.

\section*{Acknowledgment}

\par This research was financially supported by Research Council of Finland (former Academy of Finland), 6Genesis Flagship (grant no. 346208); European union’s Horizon 2020 research and innovation programme (EU-H2020), Hexa-X-II (grant no. 101095759) project; and in Brazil by CNPq (305021/2021-4, 401730/2022-0, 402378/2021-0) and RNP/MCTIC 6G Mobile Communications Systems (01245.010604/2020-14).

\bibliographystyle{./bibliography/IEEEtran}
\bibliography{./bibliography/main}

\begin{thebibliography}{10}
\providecommand{\url}[1]{#1}
\csname url@samestyle\endcsname
\providecommand{\newblock}{\relax}
\providecommand{\bibinfo}[2]{#2}
\providecommand{\BIBentrySTDinterwordspacing}{\spaceskip=0pt\relax}
\providecommand{\BIBentryALTinterwordstretchfactor}{4}
\providecommand{\BIBentryALTinterwordspacing}{\spaceskip=\fontdimen2\font plus
\BIBentryALTinterwordstretchfactor\fontdimen3\font minus \fontdimen4\font\relax}
\providecommand{\BIBforeignlanguage}[2]{{%
\expandafter\ifx\csname l@#1\endcsname\relax
\typeout{** WARNING: IEEEtran.bst: No hyphenation pattern has been}%
\typeout{** loaded for the language `#1'. Using the pattern for}%
\typeout{** the default language instead.}%
\else
\language=\csname l@#1\endcsname
\fi
#2}}
\providecommand{\BIBdecl}{\relax}
\BIBdecl

\bibitem{heath2018}
R.~W. Heath~Jr and A.~Lozano, \emph{{Foundations of MIMO Communication}}.\hskip 1em plus 0.5em minus 0.4em\relax Cambridge University Press, 2018.

\bibitem{interdonato2019}
G.~Interdonato \emph{et~al.}, ``{Ubiquitous Cell-Free Massive MIMO Communications},'' \emph{EURASIP J. Wireless Commun. Netw.}, vol. 2019, no.~1, pp. 1--13, 2019.

\bibitem{boukhedimi2019}
I.~Boukhedimi, A.~Kammoun, and M.-S. Alouini, ``{LMMSE Receivers in Uplink Massive MIMO Systems With Correlated Rician Fading},'' \emph{IEEE Trans. Commun.}, vol.~67, no.~1, pp. 230--243, 2019.

\bibitem{kassaw2018}
A.~Kassaw, D.~Hailemariam, and A.~M. Zoubirl, ``{Performance Analysis of Uplink Massive MIMO System Over Rician Fading Channel},'' in \emph{26th Eur. Signal Process. Conf. (EUSIPCO)}, 2018, pp. 1272--1276.

\bibitem{ghacham2017}
S.~Ghacham, M.~Benjillali, and Z.~Guennoun, ``{Low-complexity detection for massive MIMO systems over correlated Rician fading},'' in \emph{13th Int. Wireless Commun. Mobile Comput. Conf. (IWCMC)}, 2017, pp. 1677--1682.

\bibitem{zhang2016}
J.~Zhang \emph{et~al.}, ``{Achievable Rate of Rician Large-Scale MIMO Channels With Transceiver Hardware Impairments},'' \emph{IEEE Trans. Veh. Technol.}, vol.~65, no.~10, pp. 8800--8806, 2016.

\bibitem{berhane2017}
T.~M. Berhane \emph{et~al.}, ``{SLNR-Based Precoding for Single Cell Full-Duplex MU-MIMO Systems},'' \emph{IEEE Trans. Veh. Technol.}, vol.~66, no.~9, pp. 7877--7887, 2017.

\bibitem{he2021}
J.~He \emph{et~al.}, ``{Spectral Efficiency of the Multiway Massive System over Rician Fading Channels},'' \emph{Sec. Commun. Netw.}, vol. 2021, pp. 1--8, 2021.

\bibitem{ghacham2019}
S.~Ghacham \emph{et~al.}, ``{Rate Analysis of Uplink Massive MIMO With Low-Resolution ADCs and ZF Detectors Over Rician Fading Channels},'' \emph{IEEE Commun. Lett.}, vol.~23, no.~9, pp. 1631--1635, 2019.

\bibitem{zhang2014}
Q.~Zhang \emph{et~al.}, ``{Power Scaling of Uplink Massive MIMO Systems With Arbitrary-Rank Channel Means},'' \emph{IEEE J. Sel. Topics Signal Process.}, vol.~8, no.~5, pp. 966--981, 2014.

\bibitem{kammoun2019}
A.~Kammoun \emph{et~al.}, ``{Asymptotic Analysis of RZF in Large-Scale MU-MIMO Systems Over Rician Channels},'' \emph{IEEE Trans. Inf. Theory}, vol.~65, no.~11, pp. 7268--7286, 2019.

\bibitem{ozdogan2019}
{\"O}.~{\"O}zdogan, E.~Björnson, and E.~G. Larsson, ``{Massive MIMO With Spatially Correlated Rician Fading Channels},'' \emph{IEEE Trans. Commun.}, vol.~67, no.~5, pp. 3234--3250, 2019.

\bibitem{dileep2021}
D.~Kumar \emph{et~al.}, ``{Latency-Aware Joint Transmit Beamforming and Receive Power Splitting for SWIPT Systems},'' in \emph{IEEE 32nd Annu. Int. Symp. Personal Indoor Mobile Radio Commun. (PIMRC)}, 2021, pp. 490--494.

\bibitem{wang2012}
C.~Wang, T.~C.-K. Liu, and X.~Dong, ``{Impact of Channel Estimation Error on the Performance of Amplify-and-Forward Two-Way Relaying},'' \emph{IEEE Trans. Veh. Technol.}, vol.~61, no.~3, pp. 1197--1207, 2012.

\bibitem{eraslan2013}
E.~Eraslan, B.~Daneshrad, and C.-Y. Lou, ``{Performance Indicator for MIMO MMSE Receivers in the Presence of Channel Estimation Error},'' \emph{IEEE Wireless Commun. Lett.}, vol.~2, no.~2, pp. 211--214, 2013.

\bibitem{onel2022}
O.~L.~A. López \emph{et~al.}, ``{Massive MIMO With Radio Stripes for Indoor Wireless Energy Transfer},'' \emph{IEEE Trans. Wireless Commun.}, vol.~21, no.~9, pp. 7088--7104, 2022.

\bibitem{liu2016_2}
M.~Liu \emph{et~al.}, ``{Non-Asymptotic Outage Probability of Large-Scale MU-MIMO Systems with Linear Receivers},'' in \emph{IEEE 84th Veh. Technol. Conf. (VTC-Fall)}, 2016, pp. 1--5.

\bibitem{liu2020}
P.~Liu \emph{et~al.}, ``{Spectral Efficiency Analysis of Cell-Free Massive MIMO Systems With Zero-Forcing Detector},'' \emph{IEEE Trans. Wireless Commun.}, vol.~19, no.~2, pp. 795--807, 2020.

\bibitem{puccetti2022}
G.~Puccetti, ``{Measuring Linear Correlation Between Random Vectors},'' \emph{Inf. Sci.}, vol. 607, pp. 1328--1347, 2022.

\bibitem{mathworks_cond}
{MathWorks}, ``Condition number for inversion,'' \url{https://se.mathworks.com/help/matlab/ref/cond.html}, accessed: 2024-09-04.

\bibitem{rohde2011}
\BIBentryALTinterwordspacing
{Rohde \& Schwarz}, ``{Assessing a MIMO Channel},'' Tech. Rep., Februay 2011, accessed: 2024-09-05. [Online]. Available: \url{https://cdn.rohde-schwarz.com/pws/dl_downloads/dl_application/application_notes/1sp18/1SP18_10e.pdf}
\BIBentrySTDinterwordspacing

\bibitem{tominaga2024}
E.~N. Tominaga \emph{et~al.}, ``{Trade-Off Between Beamforming and Macro-Diversity Gains in Distributed mMIMO},'' in \emph{2024 IEEE Wireless Commun. Netw. Conf. (WCNC)}, 2024, pp. 1--6.

\end{thebibliography}

\end{document}